\documentclass{pasj00}

\begin{document}
\SetRunningHead{}{}
\Received{}
\Accepted{}

\title{A Suzaku Observation of the Low-Ionization Fe-Line Emission from RCW~86}

\author{Masaru \textsc{Ueno},\altaffilmark{1} 
        Rie \textsc{Sato},\altaffilmark{1} 
        Jun \textsc{Kataoka}, \altaffilmark{1} 
        Aya \textsc{Bamba},\altaffilmark{2} 
        Ilana \textsc{Harrus}, \altaffilmark{3,4}\\
	Junko \textsc{Hiraga},\altaffilmark{2} 
	John P.\ \textsc{Hughes},\altaffilmark{5}
        Caroline A.\ \textsc{Kilbourne}, \altaffilmark{4} 
        Katsuji \textsc{Koyama}, \altaffilmark{6} \\
	Motohide \textsc{Kokubun},\altaffilmark{7}
        Hiroshi \textsc{Nakajima}, \altaffilmark{6} 
	Masanobu \textsc{Ozaki},\altaffilmark{8}
	Robert \textsc{Petre},\altaffilmark{4}\\
	Tadayuki \textsc{Takahashi},\altaffilmark{8}
	Takaaki \textsc{Tanaka},\altaffilmark{8}
	Hiroshi \textsc{Tomida},\altaffilmark{9} and 
        Hiroya \textsc{Yamaguchi} \altaffilmark{6} 
}
\altaffiltext{1}{Department of Physics, Faculty of Science, Tokyo
Institute of Technology, \\2-12-1, Meguro-ku, Ohokayama, 
Tokyo 152-8551, Japan}
\email{E-mail(MU): masaru@hp.phys.titech.ac.jp}
\altaffiltext{2}{RIKEN (The Institute for Physics and Chemical
Research) \\2-1, Hirosawa, Wako-shi, Saitama, Japan}
\altaffiltext{3}{Department of Physics and Astronomy, Johns Hopkins
University, Baltimore, MD, 21218, USA}
\altaffiltext{4}{NASA Goddard Space Flight Center, X-ray Astrophysics Laboratory, 
Code 662, \\Greenbelt, MD 20771, USA}
\altaffiltext{5}{Department of Physics and Astronomy, Rutgers University
\\136 Frelinghuysen Road, Piscataway, NJ 08854-8109, USA}
\altaffiltext{6}{Department of Physics, Graduate School of Science,
Kyoto University, \\Sakyo-ku, Kyoto 606-8502, Japan}
\altaffiltext{7}{Department of Physics, 
University of Tokyo, 7-3-1 Hongo, Bunkyo-ku, Tokyo, Japan}
\altaffiltext{8}{Institute of Space and Astronautical Science, 
Japan Aerospace Exploration Agency \\3-1-1 Yoshinodai, Sagamihara,
Kanagawa 229-8510, Japan} 
\altaffiltext{9}{ISS Science Project Team, Institute of Space and
Astronautical Science, \\Japan Aerospace Exploration Agency 2-1-1,
Sengen, Tsukuba, Ibaraki 305-8505, Japan} 

\KeyWords{shock waves -- ISM: supernova remnants -- ISM: individual
(RCW~86) -- X-rays: ISM} 

\maketitle

\begin{abstract}
The newly operational X-ray satellite Suzaku observed the southwestern
quadrant of the supernova
remnant (SNR) RCW~86 in  February 2006 to study the nature of the 6.4~keV
emission line first detected with the Advanced Satellite for Cosmology 
and Astronomy (ASCA). The new data confirm the existence of the line,  
localizing it for the first time; most of the line emission is adjacent 
 and interior to the forward shock and not at the locus of the continuum
 hard emission. We also report the first detection of a
7.1~keV line that we interpret as the K$\beta$ emission from
low-ionization iron. The Fe-K line features are consistent with a 
non-equilibrium plasma of Fe-rich ejecta with $n_{e}t \lesssim 
10^{9}$~cm$^{-3}$~s and $kT_{e} \sim 5$~keV.  This combination of low 
$n_{e}t$ and high $kT_{e}$ suggests collisionless electron heating 
in an SNR shock. 
The Fe K$\alpha$ line shows evidence for intrinsic broadening, with a
 width of 47 (34--59)~eV (99\% error region). The difference of the
 spatial distributions of the hard continuum above 3~keV  and the Fe-K
 line emission support a synchrotron origin for the hard continuum.
\end{abstract}

\section{Introduction}

Since the first discovery, in SN~1006, of synchrotron X-ray emission 
from a supernova remnant shell \citep{koyama1995}, similar X-ray
emission characteristics have been found in a number of 
young SNRs [see for 
example, Cas~A \citep{hughes2000}; Tycho \citep{hwang2002}; or 
G347.3$-$0.5 \citep{slane2001}].  This X-ray synchrotron emission is 
usually interpreted as coming from electrons with energy up to 10 
TeV and is the strongest evidence to date of particle acceleration in 
the diffuse shocks of young SNRs.

RCW~86 is one of the SNRs in which such X-ray synchrotron emission 
has been discovered. It has in addition some high-energy emission 
associated with a low-ionization Fe-K line and localized in the
southwest part of the remnant.
This line was discovered in 1997 in the ASCA observations of the 
remnant \citep{vink1997} and confirmed  subsequently using BeppoSAX 
\citep{bocchino2000}.
The origin of both the line and the high-energy X-ray emission 
remain a mystery.
In their initial paper, \citet{vink1997} conjectured that the 
strong 6.4~keV line and the hard X-ray emission could be explained by 
an electron distribution with a supra-thermal tail. The BeppoSAX
observations revealed  spatial variations in the hardness ratio of 
the emission and favored a two-temperature plasma model, mixing a 
low-abundance, low-temperature component with a high-abundance, 
high-temperature counterpart.  The high-temperature component was
interpreted as Fe-rich 
ejecta and the low temperature component as circumstellar medium 
\citep{bocchino2000}. 
The detection of high-energy ($> 10$~keV) emission by 
RXTE \citep{petre1999} suggests the existence of non-thermal X-ray
emission.  \citet{bamba2000} and \citet{borkowski2001a} found that the
ASCA spectrum from the SW region can be explained by a combination of
three components: low and high temperature plasmas, and
non-thermal (synchrotron) emission.  Using the superior high spatial
resolution of Chandra, \citet{rho2002} spatially resolved the
low-temperature plasma and the hard continuum emission. By showing a
spatial correlation between the hard continuum and radio emission, they
provided strong evidence that the hard continuum originates from 
synchrotron emission.  Since they found the Fe K$\alpha$ line 
arises only from
the region of the hard continuum, they suggested that the
synchrotron X-rays are emitted from Fe-rich ejecta. 
Hard X-ray emission from the northeast rim of RCW~86 shows no evidence
for a 6.4~keV line, and \citet{vink2006} has suggested this emission is
synchrotron.  

Since the spatial distribution of the 6.4~keV-line and precise
spectroscopy of the line features provide important information about
the emission mechanism of the line, we performed a Suzaku observation of
the southwestern quadrant of RCW~86. In this paper, we mainly report on
the results in the hard band ($>3$~keV): the hard continuum and Fe line
features. We assume a kinematic distance of 3~kpc (\cite{rosado1996}). 

\section{SUZAKU Observation of RCW~86}

Suzaku is a joint Japanese-US mission that was launched in July 2005 
(\cite{mitsuda2006}). Two instruments, the XIS and HXD, are
operational. The XIS consists of 4 independent CCDs cameras, three
front-illuminated (energy range 0.4--12~keV) and one back-illuminated
(energy range 
0.2--12~keV).  Each is located in the focal plane of a dedicated nested
thin-foil X-ray telescope (XRTs). 
The total effective area is quite large ($\sim$
1000~cm$^{\rm -2}$ at 6~keV). In addition, the low-Earth orbit of Suzaku
provides a low, reproducible background environment. The HXD is a
non-imaging collimated detector which extends the bandpass of the
observatory up to 600~keV. More information on the mission and the 
instruments on board (XRT, XIS, and HXD) can be found in
\citet{serlemitsos2006}, \citet{koyama2006}, and \citet{takahashi2006}. 

The southwest quadrant of RCW~86 was observed by Suzaku in February
2006. At the writing of this paper, the background determination for the
HXD is still not complete and we chose not to include any HXD data in
our analysis. The XIS contamination \citep{koyama2006} plays no role at
the energies of the Fe-K line, our main concern for this paper.

\begin{table*}[th]
\begin{center}
\caption{Suzaku Observations Appearing in this Paper.}
\label{tab:obs}
  \begin{tabular}{lccc}
   \hline \hline
   Target & Observation ID & Date & Exposure$^{\rm a}$ (ks) \\
\hline
RCW~86 Southwest & 500004010 & 2006/02/12--14 & 116.5/96.4\\
North Ecliptic Polar &  500026010 & 2006/02/10--12 & 96.0/87.5 \\
High Latitude A & 500027010 & 2006/02/14--15 & 79.5/71.1 \\
Sgr~C           & 500018010 & 2006/02/20--23 & 114.5/98.6\\
\hline
  \multicolumn{4}{@{}l@{}}{\hbox to 0pt{\parbox{90mm}{\footnotesize
      \par\noindent
      \footnotemark[$^{\rm a}$]Exposure times for XIS0,2,3/XIS1 after
   the data screening.
}\hss}}
\end{tabular}
\end{center}
\end{table*}

The data (revision 0.7; \cite{mitsuda2006}) were 
analyzed after the following filtering. 
Data taken with low data rate, at low elevation angle from
the Earth rim ($<$5\arcdeg), or during passage through the South
Atlantic Anomaly were removed. We also removed data from hot and
flickering pixels using the {\tt cleansis} software. In order to
suppress the hard-band background of XIS~1, which is higher than that of 
XIS~0, 2, or 3, data taken at low cut-off rigidity regions ($<$ 6~GV)
were also removed from the XIS~1 data.  The resultant
exposure times are 116~ks for each XIS~0, 2, 3 and 96~ks for XIS~1. 
At this stage of the mission, the attitude is not completely
correct and we used a point source detected in an XMM-Newton observation 
of RCW~86, to correct the absolute coordinate system of our
observation. The correction is rather small
($\Delta$RA=$-\timeform{0.0044D}$, $\Delta$Dec=$-\timeform{0.0049D}$)
for all 4 of the XIS.

\section{XIS Images}

\begin{figure*}[ht]
\begin{center}
  \FigureFile(75mm,75mm){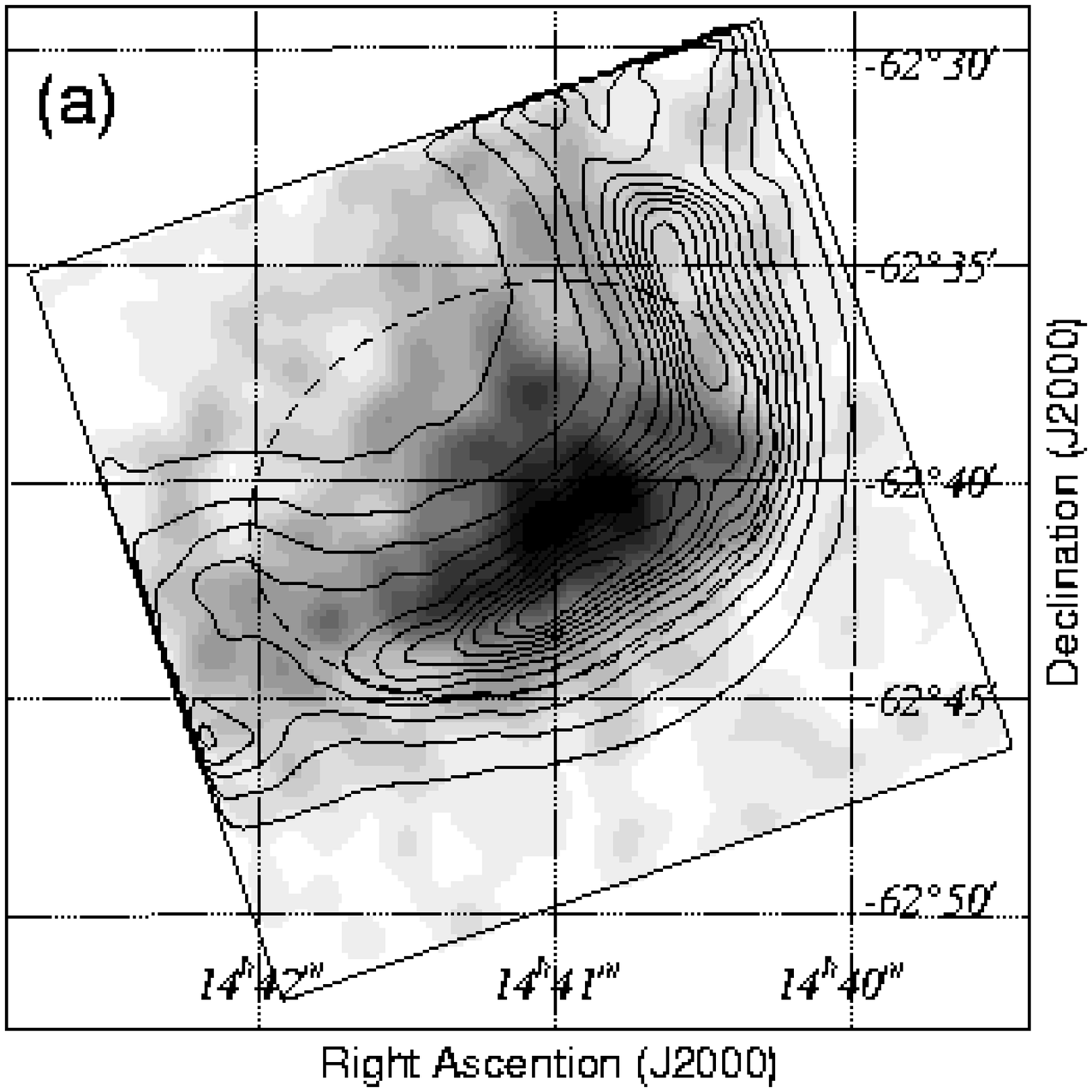}
  \FigureFile(75mm,75mm){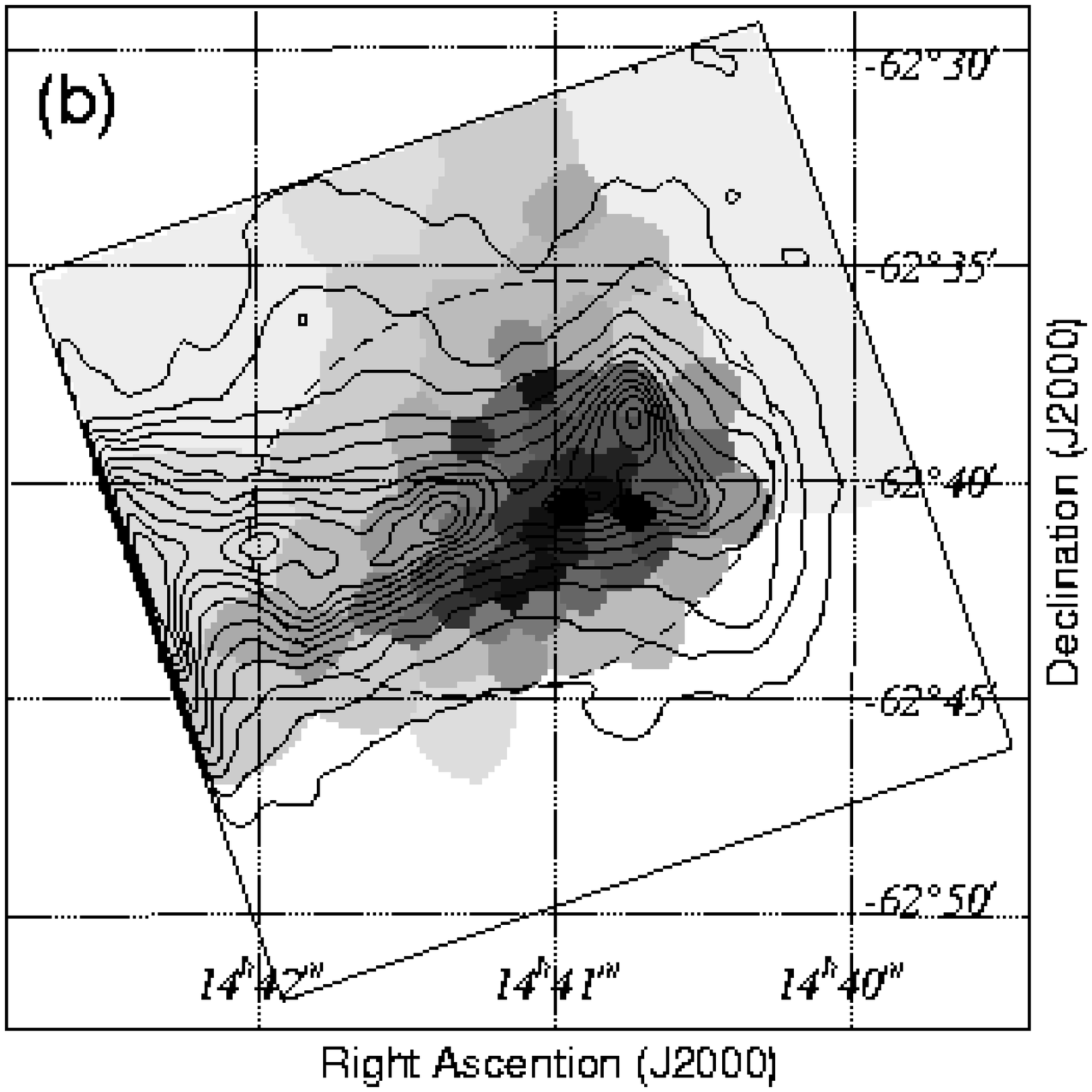}
  \caption{XIS 0.5--1.0~keV (a) and 3.0--6.0~keV (b) intensity contours
 overlaid by the 6.4~keV-line images. The 6.4~keV line image in (a) is
 smoothed with a Gaussian kernel of $\sigma$=25$\timeform{''}$, whereas
 that in (b) is  adaptively binned using the weighted Voronoi
 tessellation algorithm  \citep{diehl2006}. 
 Images and contours are plotted using a linear scale. The spectral
 region is shown with a broken-line ellipse.  The XIS field of view is
 designated with a solid-line square in each image. }
  \label{fig:2bandimages}
\end{center}
\end{figure*}

\begin{figure}[ht]
\begin{center}
  \FigureFile(80mm,80mm){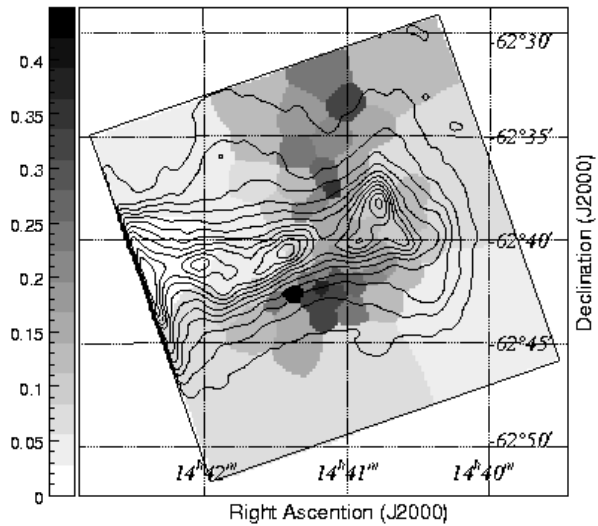}
  \caption{Count-rate ratio map between the 6.4~keV line and the
 5.0--6.2~keV band overlaid with contours of the 3.0--6.0~keV intensity
 map. An equivalent width (in keV) of the 6.4~keV line can be obtained
 by multiplying this ratio by a factor of 3.} 
  \label{fig:eqwidth}
\end{center}
\end{figure}

\begin{figure}[h]
\begin{center}
  \FigureFile(75mm,75mm){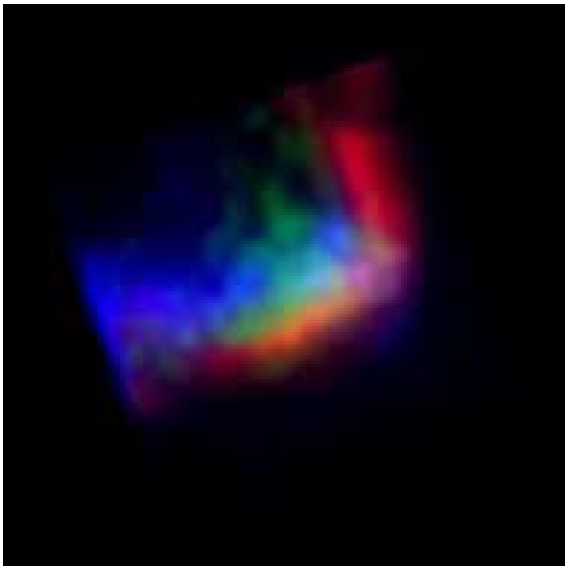}
  \caption{True color XIS image of the RCW~86 southwestern region: red
 represents 0.5--1.0~keV photons; blue represents 3.0--6.0~keV photons; 
 and green represents 6.4~keV emission.}
  \label{fig:3color}
\end{center}
\end{figure}

Figures 1a and b show XIS intensity contours in the low (0.5--1.0~keV)
and high (3.0--6.0~keV) energy bands. The high-energy band was chosen to
show the contribution from the continuum only. 
Overlaid on the low energy contours in Fig.~1a is the 6.4~keV line
intensity image. This image was generated by subtracting an  estimated
continuum level from the (6.34--6.46~keV) image. The continuum
contribution was estimated by scaling the (5.0--6.2~keV) image using a
power-law model. Since smoothing may introduce artificial structures,  
we adaptively binned the image using the weighted Voronoi tessellation 
algorithm implemented by \citet{diehl2006}, which is based on the
algorithm of \citet{cappellari2003}.  
The size of each bin was determined so that significance of the 6.4~keV 
line is at least 5$\sigma$. The resulting image is shown superposed on
the high energy contours in Fig.~\ref{fig:2bandimages}b. In order to
quantify the spatial correlation between the 6.4~keV line and the
continuum emission, we furthermore obtained the spatial variation of
count-rate ratios between the 6.4~keV line and the (5.0--6.2~keV) band
using the same binning. The result is shown in Fig.~\ref{fig:eqwidth}
with the 3.0--6.0~keV intensity contours. If we multiply this ratio by
3, it corresponds to the equivalent width (keV) of the 6.4~keV line at
that position. We have generated in Fig.~\ref{fig:3color} a true color
image of the remnant showing the spatial variations between low and
high-energy continuum (shown in blue and red respectively), and the
6.4~keV-line (shown in green).  

\begin{figure}[h]
\begin{center}
  \FigureFile(70mm,70mm){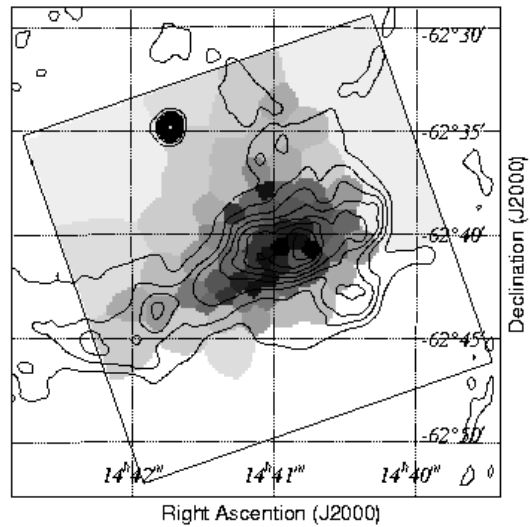}
  \caption{Intensity contours of the MOST 843~MHz radio continuum
 \citep{whiteoak1996} overlaid by the 6.4~keV-line image. Images and contours are plotted 
 using a linear scale. }
  \label{fig:fe_radio}
\end{center}
\end{figure}

The spatial distribution of the 6.4~keV line is revealed for
the first time, thanks to the combination of Suzaku's large effective
area and low background around 6.4~keV. The red region in
Fig.~\ref{fig:3color}, corresponding to the location of the
low-temperature plasma, was attributed to the forward shock by
\citet{rho2002}.  The blue region corresponds to hard, filamentary
structures revealed by the Chandra observation \citep{rho2002}. The
6.4~keV line is emitted primarily from the interior to the forward
shock region. As apparent in Fig~1b, the high-energy continuum emission 
does not correlate with the intensity peak of the 6.4~keV line.
Fig.~\ref{fig:eqwidth} shows instead that the equivalent width of the
6.4~keV line and the continuum emission anti-correlate with each other,
and thus probably have different origins. In Fig.~\ref{fig:fe_radio}, 
we demonstrate that there is good spatial correlation between the
6.4~keV line and the radio continuum emission (MOST 843MHz;
\cite{whiteoak1996}). Since radio continuum emission is often found
in the downstream of a shock, which is also pointed
out for the southwest region of RCW~86 (\cite{rho2002}), the 6.4~keV
line is likely to be emitted from the downstream of a shock. 
We will examine below how the spectral analysis of the
line fit with this new finding.  

\section{Spectral Analysis of the 6.4~keV Line}

In order to study the origin of the 6.4~keV emission line, we extracted
spectra from 
an elliptical region 
centered at (\timeform{14h41m08.7s},\timeform{-62D40'10''}),
with $\timeform{12.5'} \times \timeform{9.2'}$
major and minor axes and a position angle of 
\timeform{20D}.  This region is demarcated by
the dashed line in Fig.~\ref{fig:2bandimages}.
It was chosen to provide the maximum 
signal-to-noise ratio for the 6.4~keV line.  
While one might expect distinct spectra from the red,
green and blue 
regions in  Fig.~2, the spatial resolution of XIS does not allow such 
distinct separation. Moreover, Chandra and XMM-Newton are better 
suited for
spatially resolved spectroscopy of the soft component and the hard 
continuum emission. Therefore we concentrate here on the spectroscopy 
of the Fe-K line region. 
 
As the entire XIS field of view is covered by the remnant, we used
"blank-sky" observations for background spectra,  
extracting  the background from the same 
regions on the detectors as the ones used in the analysis. The two
observations used to provide background spectra
("North Ecliptic Polar" and "High Latitude A", see
Table~\ref{tab:obs}) were carried out immediately before and after
the RCW~86 observation, insuring an identical 
detector response.  Spectra from these two observations were weighted
averaged using their exposure times.  
Above 3~keV, differences between the responses from all the FI chips are
negligible; we therefore summed the spectra from XIS~0, 2 and 3 for the
following analysis. Although the BI spectrum is analyzed separately, the 
differences between the best fit parameters from the two spectra are
negligible.   

\begin{figure}
\begin{center}
  \FigureFile(80mm,80mm){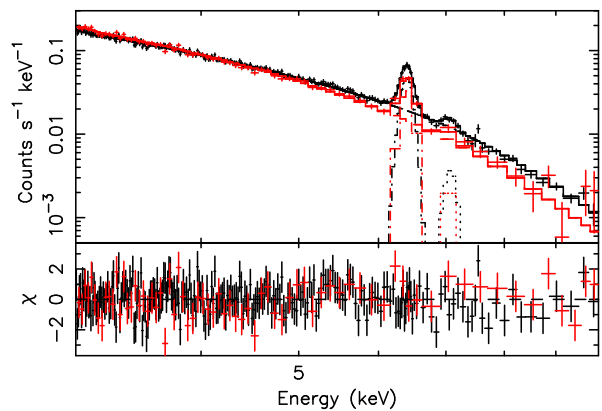}
  \caption{Background-subtracted spectra in the
 hard energy band ($>$3~keV). Data points of XIS~0+2+3 and XIS~1 are
 shown with black and red crosses, respectively.
 Solid lines show the best-fit model of a power-law function plus two
 Gaussians.} 
  \label{fig:spec}
\end{center}
\end{figure}

Background-subtracted 3.0--10~keV spectra are shown
in Fig.~\ref{fig:spec}.  Since the spectrum appears flat with the 
exception of two conspicuous lines at 6.4 and 7.1~keV, we fitted it
using a power-law model plus two Gaussians corrected for the interstellar
absorption. Since the absorption column is not well determined by these
hard band spectra,  we fixed the column density $N_{\rm H}$ at 
$5\times 10^{21}$~cm$^{-2}$, which is close to the values obtained by 
\citet{rho2002} for the SW regions. The fit results have no 
strong dependence on the column density. The standard RMF files
version 2006-02-13 were used, whereas ARF files were produced by the {\tt
xissimarfgen} software version 2006-05-28 with CALDB version 2006-05-24
assuming uniform emission from the spectral region. 
A power law plus two Gaussians yields an acceptable fit
($\chi^{2}$/degree of freedom (dof) = 270.7/291) with the best-fit
parameters given in Table~\ref{tab:spec}. The best-fit model is shown in
Figure~\ref{fig:spec}. Note that there is a systematic error of about
15~eV in the centroid energy of the lines. The 7.1~keV line is detected
clearly for the first time. 
\citet{rho2002} fitted the hard continuum component of the Chandra spectrum
with a SRCUT model \citep{reynolds1999} with  a radio index $\alpha=0.6$ and
a cut-off frequency $\nu _{c} \sim 8 \times 10^{16}$~Hz. This model has an
averaged slope of a photon index 3.0 in the 3.0--10~keV band, and the
Suzaku spectrum is almost consistent with this value.  

Although the model reproduces the overall spectrum quite well, a 
residual feature is apparent at $\sim$5.4~keV in Fig.~\ref{fig:spec}.  
Addition of a Gaussian with a fixed $\sigma$ (40~eV) improves the 
fit, with $\chi ^{2}/{\rm dof} = 258.5/289$. The best-fit
centroid energy and the flux of the Gaussian are 5.38 (5.32--5.43)~keV and 
2.7 (1.5--3.9)$\times 10^{-6}$~photons~cm$^{-2}$~s$^{-1}$,
respectively. 
The centroid energy is consistent with the 
K$\alpha$ line energy of low-ionization chromium (5.41~keV). 
We find no indication for the presence of the K$\alpha$ line of
low-ionization nickel (at 7.47~keV), obtaining a 90\% upper limit for the 
flux $F_{Ni} < 3.3 \times 10^{-6}$~photons~cm$^{-2}$~s$^{-1}$. 

In order to distinguish between the instrumental and intrinsic width of the
Fe-K lines, we compared them with the values derived from an identical
analysis for the Sgr~C observation (Table~\ref{tab:obs}) which was
performed 8 days after the RCW~86 observation. The $1\sigma$ width of
6.4~keV in the Sgr~C observation is 39 (34--44)~eV.  
Since the 6.4~keV line from Sgr~C, a molecular cloud, is expected to
have negligible intrinsic width compared with the XIS energy resolution, 
this width is thought to represent the instrumental energy resolution. 
Therefore, the difference of the 6.4~keV line 
widths between these two observations yields the intrinsic width of the
6.4~keV line from RCW~86, which is calculated to be 47 (34--59)~eV (99\%
confidence).  A similar analysis for the 7.1~keV line yields a width of 
57 ($<$115)~eV; a narrow 7.1~keV line cannot be excluded.

\begin{table}
\begin{center}
\caption{Best-fit parameters of the hard X-ray emission with a power-law
 function plus two Gaussians.}
\label{tab:spec}
  \begin{tabular}{lc}
   \hline \hline
   Parameters & Values \\
  \hline
Power-law &  \\
\hspace*{0.2cm}  $\Gamma$ \dotfill& 3.17 (3.14--3.19)\\
 \hspace*{0.2cm}  norm$^{\rm a}$\dotfill& 3.5 (3.3--3.6) \\
6.4~keV-line&\\
\hspace*{0.2cm} Center~(keV)\dotfill& 6.404 (6.400--6.407)\\
\hspace*{0.2cm} $\sigma ^{\rm b}$~(eV) \dotfill& 61 (56--66)\\
\hspace*{0.2cm} Flux$^{\rm c}$ \dotfill& 5.1 (4.9--5.3)\\
7.1~keV-line & \\
\hspace*{0.2cm} Center~(keV)\dotfill& 7.08 (7.04--7.11)\\
\hspace*{0.2cm} $\sigma ^{\rm b}$~(eV)\dotfill& 69 (35--101)\\
\hspace*{0.2cm} Flux$^{\rm c}$\dotfill& 0.53 (0.40--0.69) \\
$N_{\rm H}$ (cm$^{-2}$)\dotfill & $5\times10^{21}$ (fixed)\\
$\chi ^{2}$/dof \dotfill& 270.7/291\\ 
\hline
  \multicolumn{2}{@{}l@{}}{\hbox to 0pt{\parbox{70mm}{\footnotesize
      Notes. Error regions correspond to 90\% confidence levels. 
      \par\noindent
      \footnotemark[$^{\rm a}$]Normalization at 1 keV ($\times
   10^{-2}$~photons~keV$^{-1}$~cm$^{-2}$~s$^{-1}$). 
      \par\noindent
      \footnotemark[$^{\rm b}$]This value includes detector energy
   resolution (see text). 
      \par\noindent
      \footnotemark[$^{\rm c}$]Total flux ($\times
   10^{-5}$~photons~cm$^{-2}$~s$^{-1}$) in the line.   
    }\hss}}
\end{tabular}
\end{center}
\end{table}

\section{Discussion}

The morphology of the 6.4~keV line emission has been revealed for the
first time by the Suzaku observation, and shows that the line and the
continuum emissions have different origins, even though both arise from 
the inner part of the shell and are likely to come from ejecta. 

We first discuss the low-ionization plasma scenario for the 6.4~keV
line.  From the absence of L-shell lines \citet{borkowski2001a} have
pointed out that M-shell electrons have to be still bound. 
The Fe K$_{\beta}$ line, which was detected for the first time with
Suzaku/XIS, provides a clearer constraint on the ionization state 
of iron. The intensity ratio between Fe K$_{\alpha}$ and K$_{\beta}$
lines, which is calculated to be 0.10 (0.074--0.14) (90\% error region), 
is consistent with neutral iron, and shows that even if iron is 
ionized the mean charge is at most +12 (e.g., \cite{palmeri2003,
mendoza2004}).  The Fe K$_{\alpha}$ and K$_{\beta}$ centroid energies 
are also consistent with a low ionization 
state (e.g., \cite{palmeri2003,mendoza2004}).  
According to \citet{porquet2001}, in plasmas of
almost all temperatures found in SNRs ($5\times
10^{6}$--$8\times10^{7}$~K), iron is ionized to a mean charge of +9--+10
when the ionization parameter $n_{e}t$ is $10^{9}$~cm$^{-3}$~s.
If we interpolate the curves for $n_{e}t$= $10^{9}$ and
$10^{10}$~cm$^{-3}$~s, the mean charge of +12 corresponds to 
$n_{e}t$ $\sim 3\times 10^{9}$~cm$^{-3}$~s. 
Therefore,  $n_{e}t$ has to be $\sim 3 \times 10^{9}$~cm$^{-3}$~s or
smaller.  If we assume the plasma age 
is older than 1000~yr, the electron number density $n_{e}$ in the plasma
has to be smaller than 0.1~cm$^{-3}$.  

In order to produce the strong Fe K$\alpha$ line 
by electron inner shell ionization, there have to be a lot of electrons
which have higher energies than the Fe-K edge (7.1~keV).  
Such electrons may exist in Maxwellian or non-Maxwellian distribution. 
As a former case, we assume a NEI plasma \citep{borkowski2001b} of
$kT_{e} = 5$~keV and  $n_{e}t = 1\times10^{9}$~cm$^{-3}$~s. 
In this case, the emission measure $\int n_{e}n_{\rm Fe} dV$ 
has to be $1.4\times 10^{53}$~cm$^{-3}$. 
The elliptical extraction region of $\timeform{12.5'} \times
\timeform{9.2'}$ corresponds to $11 \times 8$~pc$^{2}$ at 3~kpc. 
Assuming a uniform density of a prolate spheroid with 
$11 \times 11 \times 8$~pc$^{3}$ axes, we arrive at $n_{e}n_{\rm
Fe} =9.1\times 10^{-6}~d_{3kpc}^{-1}$~cm$^{-6}$.  
If we apply the upper limit of the electron density, then the iron 
density $n_{Fe}$ has to be larger than $9.1 \times
10^{-5}$~d$_{3kpc}^{-1}$~cm$^{-3}$.  
This density corresponds to an iron mass of $\sim 1.3 \times
10^{32}~d_{3kpc}^{2}$~g or $\sim$0.07~M$_{\solar}$, which is feasible to
have arisen as ejecta. The largest equivalent width of the 
Fe K$\alpha$ line found in Fig.\ref{fig:eqwidth} is 1.3~keV. For an 
NEI plasma with $kT_{e} = 5$~keV and $n_{e}t=1\times 10^9$~cm$^{-3}$~s, 
this equivalent width corresponds to an iron abundance of
$\sim$2.4~solar. Since most of the detected continuum emission arises in
a different region from the Fe K$\alpha$ line, this value represents a
lower limit for the Fe abundance. Therefore, iron is likely to originate
from ejecta. 

If only collisional interaction is taking place between electrons and
ions, the equipartition between them will not be reached at $n_{e}t$
$\sim 10^{9}$~cm$^{-3}$~s \citep{itoh1984}. 
Since we can assume $T_{i} \gg T_{e}$, by substituting $Z=A=1$
(corresponding to hydrogen or proton) to equation (3) of
\citet{laming2001}, the temperature of proton $T_{i}$ is calculated to 
be 
\begin{equation}
T_{i} = 14 \left(\frac{n_{e}t}{1\times 10^{9}~{\rm
			    cm^{-3}}}\right)^{-1}
\left(\frac{kT_{e}}{5~{\rm keV}}\right)^{\frac{5}{2}}~{\rm MeV}. 
\end{equation}
This proton temperature corresponds to a shock speed of
$8\times10^{4}$~km~s$^{-1}$, which is too high for a shock in an
SNR. Therefore, protons should have a significantly lower temperature 
and electrons have to be heated by some process other than Coulomb
collisions. The most plausible mechanism is 
collisionless heating in SNR shocks (see \cite{laming2000}). 
If electrons are heated by collisionless mechanism, the distribution of 
electrons can be non-Maxwellian or non-thermal function. 
However, since no continuum X-ray emission associated with the 6.4~keV
line is detected, we can give no observational constraint on the distribution
function of electrons. 

The Suzaku observation shows that hard X-ray continuum interior to
the shell is not related to the 6.4~keV line, and therefore likely to be 
synchrotron emission.  In order to meet the conditions for synchrotron 
X-ray emission under the standard diffusive shock acceleration model
(e.g., \cite{drury1983}), the shock velocity has to be larger
than $\sim 2000$~km~s$^{-1}$ independent of the magnetic field strength
(\cite{aharonian1999}).   This value is discrepant with the blast-wave 
velocity of $\sim$400--900~km~s$^{-1}$ determined from observations of
Balmer-dominated shocks \citep{ghavamian2001}. It is plausible that
shocks in the inner part of the shell may have higher velocity than the
forward shock. The intrinsic width of $\sim 47$~eV found for the 6.4~keV
line may be evidence of this high velocity.   
The forward shock might have decelerated by colliding with a cavity wall, 
while the inner shock relatively undecelerated within the
cavity. Alternatively, a reflected shock after the collision between the
forward shock and the interstellar medium structure to the southwest may
be propagating back into the cavity. A cavity wall interaction of RCW~86
has also been suggested to explain the overall morphology 
\citep{vink1997} and the synchrotron X-ray emission in the north-east
region \citep{vink2006}.   
Finally, in this complex region, it is possible that projection
effects cause an apparent superposition of an interacting shock and a 
much faster, relatively unimpeded one.

Finally, an observation with higher energy resolution of the SW region
is highly encouraged. A stricter constraint on the iron ionization state
can be obtained by measuring the difference between the centroid
energies of the 6.4 and 7.1~keV lines at several eV resolution
(e.g.,~\cite{palmeri2003,mendoza2004}). The intrinsic width of the
6.4~keV line and existence of Cr-K line can also be verified clearly. 

\section{Summary}

With the large effective area and low background of XIS in the hard
X-ray band, we obtained following observational results regarding the hard
X-ray continuum and Fe-line features from the SW region of RCW~86.   

\begin{enumerate}
\item The spatial distribution of the 6.4~keV line is revealed for the
      first time.  The line emission has a distinct morphology from that
      of the hard X-ray  continuum emission.
\item The Fe K$\beta$ line is detected for the first time. 
\item The Fe K$\alpha$ line is intrinsically
      broadened with a width of about 50~eV. 
\end{enumerate} 

These results are all consistent with a model in 
which the 6.4~keV line is emitted
from a low-ionization plasma of Fe-rich ejecta. 
The different morphology between the hard continuum 
6.4~keV line emission supports the interpretation of the
hard continuum as synchrotron emission.

\bigskip
We thank all the $Suzaku$ members. M.U., R.S., H.N., and H.Y. are
supported by JSPS Research Fellowship for Young Scientists. JPH
acknowledges support by NASA grant No.\ NNG05GP87G.


\end{document}